\documentclass{article}
\usepackage{spconf,amsmath,graphicx}

\usepackage{url,times,booktabs, tabularx}
\usepackage{multirow}
\usepackage{caption}
\usepackage{subcaption}
\usepackage{enumitem}
\usepackage{color}

\makeatletter
\newcommand{\thickhline}{%
    \noalign {\ifnum 0=`}\fi \hrule height 1pt
    \futurelet \reserved@a \@xhline
}
\newcolumntype{"}{@{\hskip\tabcolsep\vrule width 1pt\hskip\tabcolsep}}
\makeatother

\title{A GENERAL NETWORK ARCHITECTURE FOR SOUND EVENT LOCALIZATION AND DETECTION USING TRANSFER LEARNING AND RECURRENT NEURAL NETWORK}
\name{\begin{tabular}{c}Thi Ngoc Tho Nguyen$^{\star}$, Ngoc Khanh Nguyen, Huy Phan$^{\ddagger}$, Lam Pham $^{\diamond}$, \\ Kenneth Ooi$^{\star}$, Douglas L. Jones$^{\dagger}$, Woon-Seng Gan$^{\star}$
\sthanks{\scriptsize{This research was conducted at Singtel Cognitive and Artificial Intelligence Lab for Enterprises (SCALE@NTU), which is a collaboration between Singapore Telecommunications Limited (Singtel) and Nanyang Technological University (NTU) that is funded by the Singapore Government through the Industry Alignment Fund ‐ Industry Collaboration Projects Grant.}}
\end{tabular}
}

\address{
$^{\star}$ School of EEE, Nanyang Technological University, Singapore\\
$^{\ddagger}$ School of EECS, Queen Mary University of London, UK \\
$^{\diamond}$ School of Computing, University of Kent, UK \\
$^{\dagger}$ Department of ECE, University of Illinois at Urbana-Champaign, USA 
}

\begin{document}
\ninept
\maketitle
\begin{abstract}
Polyphonic sound event detection and localization (SELD) task is challenging because it is difficult to jointly optimize sound event detection (SED) and direction-of-arrival (DOA) estimation in the same network. 
We propose a general network architecture for SELD in which the SELD network comprises sub-networks that are pre-trained to solve SED and DOA estimation independently, and a recurrent layer that combines the SED and DOA estimation outputs into SELD outputs. The recurrent layer does the alignment between the sound classes and DOAs of sound events while being unaware of how these outputs are produced by the upstream SED and DOA estimation algorithms. This simple network architecture is compatible with different existing SED and DOA estimation algorithms. It is highly practical since the sub-networks can be improved independently. 
The experimental results using the DCASE 2020 SELD dataset show that the performances of our proposed network architecture using different SED and DOA estimation algorithms and different audio formats are competitive with other state-of-the-art SELD algorithms. The source code for the proposed SELD network architecture is available at Github\footnote{https://github.com/thomeou/General-network-architecture-for-sound-event-localization-and-detection}.

\end{abstract}
\section{Introduction}
\label{sec:intro}
Polyphonic sound event localization and detection (SELD) find a wide range of applications in urban sound sensing~\cite{Salamon2017Cnn}, wild life monitoring~\cite{Stowell2016bird}, surveillance~\cite{Foggia2016Surveillance}, autonomous driving~\cite{nandwana2016car}, and robotics~\cite{valin2004localization}. The SELD task~\cite{Adavanne2019seld} recognizes the sound class, and estimates the direction-of-arrival (DOA), the onset, and offset of a detected sound event. Polyphonic SELD refers to cases where multiple sound events overlap in time. 

SELD is an emerging topic in audio processing. It consists of two subtasks, which are sound event detection (SED) and DOA estimation (DOAE). These two subtasks are mature research topics, and there exists a large body of effective algorithms for SED and DOAE~\cite{cakir2017convolutional, mohan2008localization}. Over the past few years, majority of the methods proposed for SELD have focused on jointly optimizing SED and DOAE in the same network. Hirvonen formulated the SELD task as multi-class classification where the number of output classes is equal to the number of DOAs times the number of sound classes~\cite{hirvonen2015classification}. 
Adavanne \emph{et al.} proposed a single-input multiple-output CRNN model called SELDnet that jointly detects sound events and estimates their DOAs~\cite{Adavanne2019seld}. 
It has been shown that the joint optimization indeed affects the performance of both the SED and DOAE subtasks. Alternatively, Cao \emph{et al.} proposed a two-stage strategy to train two separate SED and DOA models~\cite{cao2019polyphonic} and used SED outputs as masks to select DOA outputs. This training scheme significantly improves the SELD performance over the jointly-trained SELDnet. Cao \emph{et al.} later proposed a jointly-trained SELD network~\cite{cao2020joint} that takes raw audio waveform as input and segregates the SELD output into event-independent tracks of events, which was first proposed in \cite{tho2020smn}. 
Huy \emph{et al.} improved jointly-trained SELD models by adding an attention layer and using mean-square-error loss for SED instead of cross-entropy loss~\cite{huy2020loss}. The top-ranked solution for DCASE 2020 SELD challenge improved the jointly-trained models by synthesizing a larger dataset from the provided data and exploiting a large ensemble of complex networks~\cite{du2020dcasetop}.      

The advantage of a jointly-trained network for SELD is clear since it needs only one joint network and requires a single forward-pass on the audio signal to produce the final predictions. However, it is challenging to jointly train multi-task networks as large models are prone to over-fitting and the subtasks' convergence may be out-of-sync ~\cite{wang2020multitask}. In the context of SELD, when the SED and DOAE subtasks share a sub-network, this sub-network is potentially pulled in different directions during the joint optimization because the former relies on spectro-temporal patterns of the audio signals while the latter relies on the phase or magnitude differences between the input channels. In addition, joint training requires datasets with joint annotations. However, such datasets are difficult to collect and annotate accurately. The current most popular SELD dataset was simulated and limited to 10-hour long~\cite{politis2020SELDTdataset}. 

\begin{figure}[tb]
\centering
\includegraphics[width=0.4\textwidth]{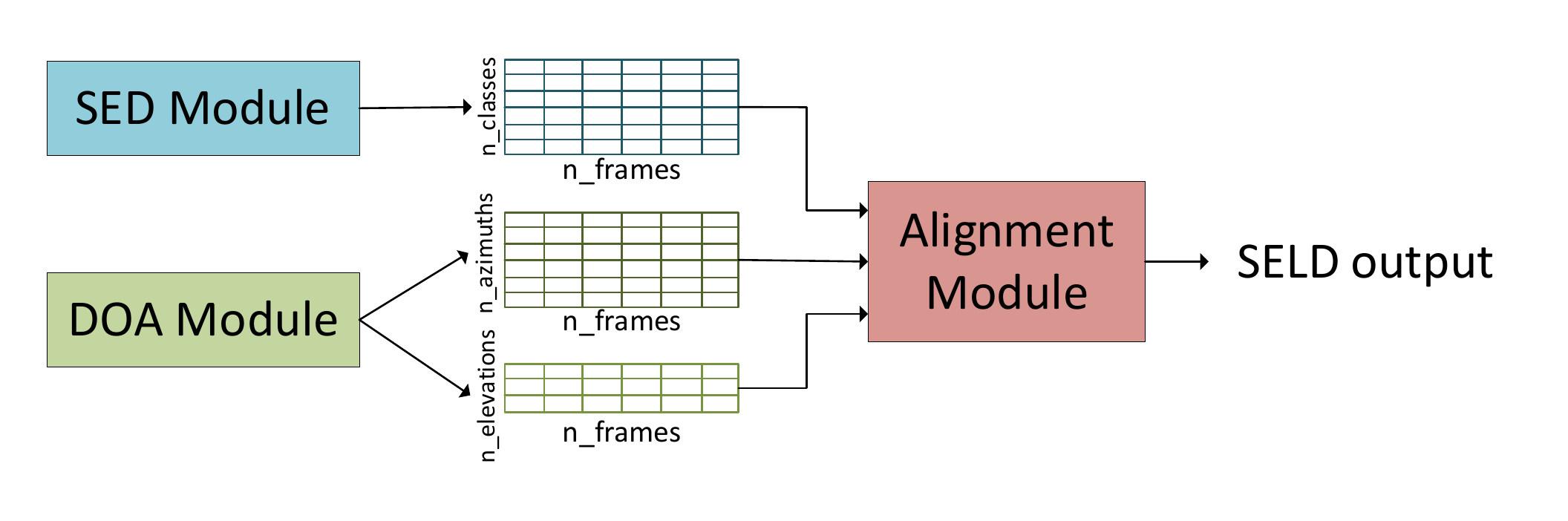}
\captionsetup{belowskip=0pt}
\vspace{-0.35cm}
\caption{A general SELD network architecture.}
\label{fig1:seld_framework}
\vspace{-0.25cm}
\end{figure} 

Our previously proposed sequence matching network (SMN) shows that it might be more beneficial to train SED and DOAE separately than jointly~\cite{tho2020smn, tho2020smn_ensemble}. However, these SMNs are tied to a signal processing-based method for DOAE and it is not straightforward to accommodate other DOAE algorithms. In this paper, we propose a novel network architecture for SELD as shown in Fig.~\ref{fig1:seld_framework}. In this architecture, the networks for SED and DOAE are pre-trained independently. An alignment network based on recurrent neural network (RNN) is then trained to align the SED and DOA output sequences on the basis that overlapping sounds often have different onsets and offsets. By matching the onsets, the offsets, and the active segments in the output sequences of the sound event detector and the DOA estimator, we can associate the estimated DOAs with the corresponding sound classes. 
For 2D SELD, the DOAE module only needs to estimate azimuth. When 3D SELD is required, both azimuth and elevation will be separately estimated by the DOAE module. The azimuth and elevation decoupling significantly reduces the dimension of the DOA outputs. We choose classification format for both SED and DOAE subtasks since it is generally easier to optimize a classification model than a regression model. In addition, the classification is necessarily multi-label to tackle the polyphonic events. The SELD output can be in class-wise format~\cite{Adavanne2019seld, cao2019polyphonic} or track-wise format~\cite{tho2020smn, cao2020joint}. 
The proposed SELD network architecture offers several advantages. First, it is easier to optimize as the SED and DOAE modules are trained separately. Second, as a generic framework, it can accommodate various SED and DOAE algorithms which may be constrained to a specific application. Third, the network architecture offers a robust SELD system without unwanted association between sound classes and DOAs since the SED and DOAE modules are pre-trained independently. Fourth and most importantly, the proposed network architecture is highly practical. Each module can be improved independently and existing task-specific SED or DOAE datasets (rather than joint annotation) can be utilized for fine-tuning. In addition, we argue that the alignment network requires smaller joint SELD datasets to train compared to a joint SELD model trained from scratch since the alignment network is light-weight, uses high-level inputs, and does not need to know how SED and DOAE outputs are produced by the upstream modules. 

In this paper, we demonstrate the practicality and efficacy of the proposed architecture by incorporating different SED and DOAE algorithms for both first-order ambisonic (FOA) and mic-array format. Specially, for each input format, we pre-train two different SED models and two different DOA models. Transfer learning is used for one SED model to demonstrate how available datasets can be utilized for SED. One of the DOA model is based on signal-processing algorithm while the other purely relies on deep learning. A bidirectional GRU is used to realize the RNN in the alignment network. Experimental results using the DCASE 2020 SELD dataset show that our proposed framework obtains competitive performances compared to other state-of-the-art SELD algorithms. The rest of our paper is organized as follows. Section II describes our proposed SELD framework. Section III presents the experimental results and discussions. Finally, we conclude the paper in Section IV.

\section{A general network architecture for SELD}
\label{sec:seld_framework}

Fig.~\ref{fig1:seld_framework} shows the block diagram of the proposed SELD framework. Both the SED and DOAE subtasks are formulated as multi-label multi-class classification to tackle the multiple-source problem. Each module takes in its respective input features and produces classification outputs for each frame. Particularly. the DOAE module has two output branches for azimuth and elevation. 
In the case of 2D SELD, the elevation branch can be removed. The SED and DOAE outputs are then concatenated along the classification-output dimension and presented to the alignment network whose task is to associate the sound classes and the DOAs. The alignment network, which is implemented by an RNN, is learned to produce SELD predictions either in class-wise or track-wise format. We use class-wise format here for simplicity purpose. The whole SELD system is trained in two stages. First, the SED and DOAE modules are pre-trained independently. After that, the alignment network is trained by treating the SED and DOAE modules as feature extractors. The weights of SED and DOAE modules in the second training stage can either be fixed or fine-tuned. In this work, we fix the weights of the pre-trained SED and DOAE models. 

\begin{figure*}[t]
\centering
\includegraphics[width=.9\textwidth]{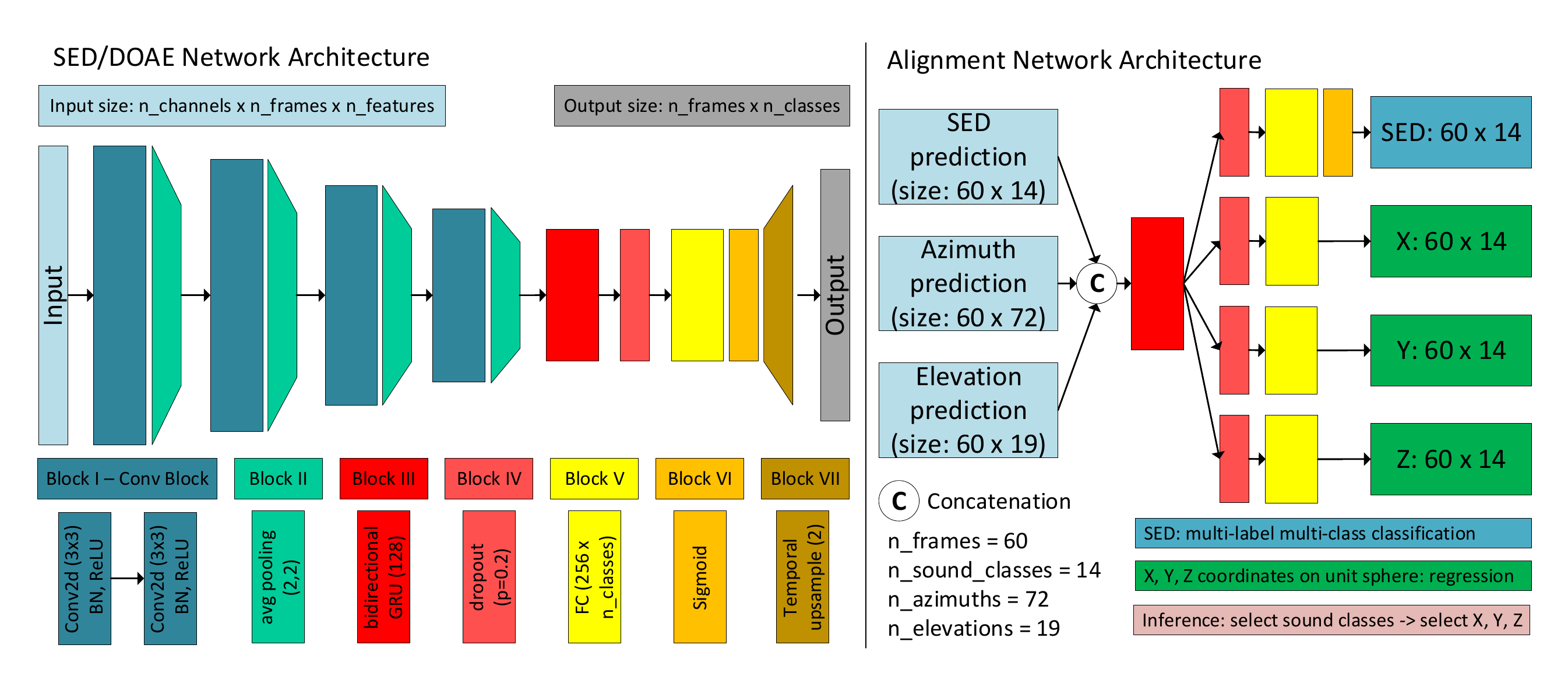}
\vspace{-0.4cm}
\caption{Left: Block diagram of the SED and DOAE networks.  $n_{frames}$ is the number of input frames. $n_{channels}$ and $n_{features}$ depend on types of input features. Both SED and DOAE are formulated as multi-label multi-class classification. Right: Block diagram of the alignment network. During inference, classes with prediction probabilities above a SED threshold are selected as active classes. DOAs of the events associated with these classes are selected correspondingly. $n_{frames}$ in the alignment network is different from the $n_{frames}$ in the SED/DOAE network due to difference in input feature frame rate and label frame rate in the DCASE 2020 dataset.}
\label{fig2:crnn_block}
\end{figure*}  

\subsection{Sound event detection}

Both SED and DOAE modules are built using a convolutional recurrent neural network (CRNN) as illustrated in Fig.~\ref{fig2:crnn_block}. The CRNN architecture consists of $4$ Conv blocks, followed by bidirectional gated recurrent units (GRUs) of size $128$, and a fully connected (FC) layer. The SED network has $1$ layer of GRU with hidden size of $128$. The numbers of filters of the $4$ Conv blocks are shown in Table~\ref{table1:subnets}. For each audio format, we train two SED models that use multi-channel and single-channel log-Mel spectrogram with $128$ and $64$ filters as inputs, respectively. To demonstrate the flexibility of the SELD framework, we train another SED model, SED-T, using transfer learning. The weights of the $4$ Conv blocks of the SED-T model is initialized using a pre-trained convolutional neural network (CNN) model named $\emph{Cnn14\_mAP=0.431}$~\cite{kong2019panns}, which was trained on the Audioset dataset~\cite{gemmeke2017audioset}. Mix-up, frequency shift, random-cutout, and specaugment are used for data augmentation~\cite{tho2020smn_ensemble}. Both of the SED models are trained using binary cross-entropy loss. 

\begin{table*}\small
\centering
\caption {Hyper-parameters for SED and DOAE networks}  
\vspace*{1pt}
\label{table1:subnets}
\scalebox{0.75}{
\begin{tabular}{r c c c c c c c c }
\hline 
Model & Audio format & Output & Input feature & \# of input channel & \# of input features & \# of Conv2d filters & \# of GRU layer & GRU hidden size \\ 
\hline 
SED-M & FOA, mic-array & n\_sound\_classes=14 & mutil-channel log-mel & 4 & 128 (Mel filters) & 64-128-256-512 & 1 & 128 \\ 
SED-T & FOA, mic-array & n\_sound\_classes=14 & single-channel log-mel & 1 & 64 (Mel filters)& 64-128-256-512 & 1 & 128 \\ 
DOA-iv & FOA & n\_azis=72, n\_eles=19 & intensity vector & 3 & 128 (Mel filters) & 32-64-128-256 & 2 & 128 \\ 
DOA-gcc & mic-array & n\_azis=72, n\_eles=19 & GCC-PHAT & 6 & 128 (time lags) & 32-64-128-256 & 2 & 128 \\ 
AZI-hist & FOA & n\_azis=72 & azimuth histogram & 1 & 72 (n\_azis) & 32-64-128-256 & 2 & 128 \\ 
ELE-hist & FOA & n\_eles=19 & elevation histogram & 1 & 19 (n\_eles) & 16-32-64-128 & 2 & 64 \\ 
\hline 
\end{tabular} 
}
\end{table*}

\subsection{Direction-of-arrival estimation}

We train a multi-task multi-label CRNN model that predicts azimuth and elevation separately. The DOAE network has $4$ Conv blocks, followed by $2$ bidirectional GRUs of size $128$ and $2$ FC layers. One FC layer outputs azimuth estimation, and the other outputs elevation estimation. Details of the DOAE networks are shown in Table~\ref{table1:subnets}. The input features of the DOA-iv and DOA-gcc models for FOA and mic-array format are intensity vectors (IV) and generalized cross-correlation with phase transform (GCC-PHAT), respectively. 

To demonstrate that the SELD network architecture can also be used with signal processing-based DOAE methods, we use a single-source (SS) histogram algorithm that was used in our previous proposed SMN~\cite{tho2020smn, tho2020smn_ensemble}. This algorithm outputs a directional histogram of SS bins for each input frame. More information about this method can be found in \cite{tho2020smn_ensemble} and \cite{tho2020taslp}. To convert the histogram into the multi-label multi-class classification format, we marginalize the 2D histograms into two 1D histograms of azimuth and elevation for each frame. Then, we stack the 1D azimuth histograms of consecutive frames together and use these 2D pseudo images as input features for a CRNN model (AZI-hist) to predict azimuth. Similar procedure is used for elevation model ELE-hist. The details of the AZI-hist and ELE-hist networks are shown in Table~\ref{table1:subnets}. Note that we only use the FOA format to train the AZI-hist and ELE-hist since the provided steering vector of the mic-array format is convoluted and not directly applicable for the SS histogram method.

\subsection{Alignment network using RNN}

The core component of the alignment network is an RNN as shown in Fig.~\ref{fig2:crnn_block}. As previously mentioned, we realize the RNN using two bidirectional GRU layers. The hidden size of the GRU is $128$. In this paper, we use the class-wise output format for SELD to simplify the optimization process. This proposed alignment network can be easily modified to suit different SELD output formats by changing the FC layers and their corresponding activation layers~\cite{tho2020smn}. SED is formulated as multi-label multi-class classification while DOAE is formulated as regression. The total loss of the alignment network is a weighted sum of SED binary cross-entropy loss and DOA regression's mean-squared error loss. We only computed DOA mean-squared error loss for frames with labelled active classes. For each frame, the alignment network outputs the probabilities of all sound classes, and their DOAs. The DOA output format is the $(x, y, z)$ coordinate on the unit sphere. During inference, we first select active classes whose classification probabilities are above a SED threshold. After that, the the DOA values corresponding to these active events are selected.

\section{Experimental results and discussions}
\label{sec:majhead}

We used the DCASE 2020 SELD dataset~\cite{politis2020SELDTdataset} for our experiments. This dataset provides both FOA and mic-array format with $4$ microphones. The dataset consists of $400$, $100$, and $100$ one-minute audio clips for training, validation, and testing, respectively. There are $14$ sound classes. 
The azimuth and elevation ranges are $[-180^{\circ}, 180^{\circ})$ and $[-45^{\circ}, 45^{\circ}]$, respectively. We used an angular resolution of $5^{\circ}$. As a result, the number of discrete azimuths and elevations was $n_{azimuths} = 72$ and $n_{elevations} = 19$ respectively. 
Validation set was used for model selection while test set was used for evaluation.  

\begin{table*}\small
\centering
\caption {Experimental results for SELD. $\downarrow$ indicates the lower the better. $\uparrow$ indicates the higher the better.}  
\label{table2: test_result}
\scalebox{0.8}{
\begin{tabular}{|c|c|c|c|c|c|c|c|c|c|c|c|c|c|}
\cline{5-14} 
\multicolumn{4}{c|}{} & \multicolumn{5}{c|}{FOA} & \multicolumn{5}{c|}{Mic-array} \\ 
\hline 
Group & SELD model & SED model & DOA model & ER$\downarrow$ & F1(\%)$\uparrow$ & DE$\downarrow$ & FR(\%)$\uparrow$ & SELD$\downarrow$ & ER$\downarrow$ & F1(\%)$\uparrow$ & DE$\downarrow$ & FR(\%)$\uparrow$ & SELD$\downarrow$ \\ 
\thickhline 
\multirow{4}{*}{\parbox{1.2cm}{\centering Baseline}}
& SELDnet~\cite{politis2020SELDTdataset} & - & - & 0.720 & 37.4 & 22.8$^{\circ}$ & 60.7 & 0466 & 0.780 & 31.4 & 27.3$^{\circ}$ & 59.0 & 0.506 \\ 
\cline{2-14}
& SELD-Huy~\cite{huy2020loss} & - & - & 0.600 & 49.2 & 19.0$^{\circ}$ & 65.6 & 0.390 & 0.590 & 50.8 & 18.2$^{\circ}$ & 64.1 & 0.380 \\ 
\cline{2-14} 
& SELD-Cao~\cite{cao2020joint} & - & - & 0.470 & 61.5 & 16.7 & 75.4 & 0.298 & - & - & - & - & - \\ 
\cline{2-14} 
& SMN~\cite{tho2020smn_ensemble} & - & - & \bf{0.401} & \bf{66.6} & \bf{15.0$^{\circ}$} & 81.0 & \bf{0.252} & - & - & - & - & - \\ 
\thickhline 
\multirow{4}{*}{\parbox{1.2cm}{\centering Proposed SELD framework}}
& SELD-M & SED-M & DOA-iv or DOA-gcc & 0.445 & 62.5 & 19.7$^{\circ}$ & 80.5 & 0.281 & 0.478 & 59.2 & \bf{23.6$^{\circ}$} & 78.8 & 0.307 \\ 
\cline{2-14}  
& SELD-T & SED-T & DOA-iv or DOA-gcc & 0.424 & 64.1 & 18.2$^{\circ}$ & 80.1 & 0.267 & \bf{0.455} & \bf{61.0} & 23.8$^{\circ}$ & \bf{81.7} & \bf{0.290} \\ 
\cline{2-14} 
& SELD-M-hist & SED-M & AZI-hist + ELE-hist & 0.443 & 62.8 & 18.0$^{\circ}$ & 80.3 & 0.279 & - & - & - & - & - \\ 
\cline{2-14} 
& SELD-T-hist & SED-T & AZI-hist + ELE-hist & 0.437 & 63.0 & 19.6$^{\circ}$ & \bf{82.2} & 0.276 & - & - & - & - & - \\ 
\hline 
\end{tabular} 
}
\end{table*}

\subsection{Evaluation metrics}

The 2020 SELD evaluation metrics~\cite{mesaros2019seldeval}, which are the official metrics of the DCASE 2020 SELD challenge, were used to evaluate the SELD performance. 
A sound event was considered a correct detection if it has correct class prediction and its estimated DOA is less than $20^{\circ}$ from the DOA ground truth. The DOA metrics were computed for each class before averaging across all classes. The DCASE 2020 SELD task adopted four evaluation metrics: DOA-dependent error rate (ER), F1-score for SED; and SED-dependent DOA error (DE), frame recall (FR) for DOA. A good SELD system should have lower ER, high F1, low DE, and high FR. We also reported SELD error which was computed as $SELD = (ER + (1-F1) + DE/180 + (1-FR))/4$ to aggregate all four metrics. In addition, we used segment-based ER and F1 to evaluate the SED networks with segment length of $1$ second. We used mean average precision to evaluate azimuth and elevation classification to avoid the usage of a threshold. 

\subsection{Hyper-parameters and training procedure}

Hyper-parameters for audio processing are sampling rate of $24$ kHz, window length of $1024$ samples, hop length of $300$ samples ($12.5$ ms), Hann window, and $1024$ FFT points. As a result, the input frame rate for SED and DOAE networks was $80$ frames per second. The input and output frame rate of the alignment network was the same as the label frame rate, which was $10$ frames per second. Because the SED and DOAE networks made use of pooling four times with a kernel size of $(2,2)$, we temporally up-sampled the outputs of these networks by a factor of $2$ to match the label frame rate.  We used inputs of length $4$ seconds to train SED and DOAE models, and input lengths of $6$ seconds to train the alignment networks. The loss weights for SED and DOAE in the alignment network were set to $(0.7, 0.3)$. Adam optimizer was used to train all the models. Learning rate was set to $0.001$ and gradually decreased to $0.0001$. The SED-T models with transferred weights were fine-tuned for $20$ epochs. The number of training epochs for the SED/DOAE and the alignment network were $60$ and $100$, respectively. A threshold of $0.3$ was used to decide active classes in the SED outputs.

\subsection{Baselines and the proposed SELD models}

We mixed and matched different pre-trained SED and DOA models with the alignment models to form different SELD models as shown in Table~\ref{table2: test_result}. We compared these SELD models with top-ranked SELD systems in the DCASE 2020 SELD challenge. We selected baselines that used only one audio format and did not use ensemble for a fair comparison. The following four baselines were considered:  

\begin{itemize}[leftmargin=*]
\item \textbf{SELDnet}: jointly-trained SELD model~\cite{politis2020SELDTdataset}, baseline of DCASE 2020 SELD challenge,
\item \textbf{SELD-Huy}: jointly-trained SELD model with attention and MSE loss for both SED and DOAE~\cite{huy2020loss}, ranked $6^{th}$,
\item \textbf{SELD-Cao}: jointly-trained SELD model with track-wise output format~\cite{cao2020joint}, ranked $4^{th}$,
\item \textbf{SMN}: our previously-proposed SMN for SELD~\cite{tho2020smn_ensemble}, whose an ensemble ranked $2^{nd}$. 	
\end{itemize}

\subsection{SELD experimental results}

\begin{table}\small
\centering
\caption {Experimental results for SED}  \vspace*{1pt}
\label{table3: sed_result}
\scalebox{0.8}{
\begin{tabular}{|c|c|c|c|c|}
\cline{2-5}
\multicolumn{1}{c|}{} & \multicolumn{2}{c|}{FOA} & \multicolumn{2}{c|}{Mic-array} \\ 
\hline 
Model & ER & F1 & ER & F1 \\ 
\hline 
SED-M & 0.302 & 80.0 & 0.302 & 79.2 \\ 
\hline 
SED-T & \bf{0.263} & \bf{82.1} & \bf{0.266} & \bf{82.1} \\ 
\hline 
\end{tabular} 
}
\end{table}

\begin{table}\small
\centering
\caption {Experimental results for DOAE}  \vspace*{1pt}
\label{table4: doa_result}
\scalebox{0.8}{
\begin{tabular}{|c|c|c|c|c|}
\hline 
Model & DOA-iv & DOA-gcc & AZI-hist & ELE-hist \\ 
\hline 
Azimuth mAP & 0.339 & 0.426 & \bf{0.509} & - \\ 
\hline 
Elevation mAP & 0.391 & \bf{0.406} & - & 0.400 \\ 
\hline 
\end{tabular} 
}
\end{table}

Table~\ref{table3: sed_result} shows the experimental results for SED on the test set. For both FOA and mic-array audio formats, the model SED-T with transferred weights outperforms the model SED-M which was trained from scratch. This confirms the benefit of transfer learning in improving the performance of the SED models. Tabel~\ref{table4: doa_result} shows the experimental results for DOAE. The AZI-hist model obtained the best mAP score for azimuth. The SS histogram method was developed to tackle multi-source cases in reverberant and noisy environments by using only SS time-frequency bins to estimate DOA. As expected, it performs better than DOA-gcc and DOA-iv which were trained from GCC-PHAT and IV features without any treatment to deal with multi-source, reverberation and noise. Elevation is more difficult to estimate than azimuth. All three models have similar mAP scores for elevation estimation. 

Table~\ref{table2: test_result} shows the experimental results for the joint task SELD. All the proposed SELD models result in very competitive performance compared to the baseline models. The SELD models for mic-array format have lower performance than the SELD models for FOA format. For FOA format, the model SED-T is ranked second just after the SMN. For mic-array format, the model SED-T is ranked first thanks to the absence of baseline models in mic-array format. These results show that our proposed network architecture work well with different audio formats and different sub-networks for both SED and DOAE. Even though the DOA-iv and DOA-gcc models result in lower standalone performance than the AZI-hist and ELE-hist, the joint SELD models formed by the DOA-iv and DOA-gcc models achieve similar performance as the joint SELD model formed by AZI-hist and ELE-hist. The SELD models that use the SED-T model with transfer learning from the Audioset dataset performs slightly better than the SELD models that use the SED-M. These performance gains could not be obtained if multi-channel input features were used to train the SED models because we do not have any large-scale multi-channel dataset available at the moment. 
The performance of our proposed SELD models is lower than those of the SMN model for FOA format most likely because the SMN combines part of the DOAE network and the alignment network. This results suggest that we should fine-tune the whole SELD models after pre-training the sub-networks for SED and DOAE. 

\section{Conclusions}

In conclusion, we have proposed a simple yet effective network architecture for SELD with pre-trained sub-networks for SED and DOAE, and an RNN-based alignment network that matches and fuses the SED and DOAE outputs. For future work, we would like to explore if fine-tuning the pre-trained SED and DOAE components could further improve the SELD performance. There is also room for improvement in terms of architectures for the alignment network.




\bibliographystyle{IEEEbib}
\bibliography{refs}

\end{document}